\documentclass{ws-procs9x6}

\usepackage{amsmath}
\usepackage{amssymb}
\usepackage{hyperref}

\newcommand{\exclude}[1]{}

\newcommand{\beq}{\begin{equation}}
\newcommand{\eeq}{\end{equation}}
\newcommand{\bea}{\begin{eqnarray}}
\newcommand{\eea}{\end{eqnarray}}

\def\la{\langle }
\def\ra{ \rangle }
 \newcommand{\junk}[1]{}

\begin{document}

\title{ Deconfinement Phase Transition in Hot and Dense  QCD at  Large N}

\author{Ariel R. Zhitnitsky}

\address{Department of Physics and Astronomy, \\
University of
  British Columbia, \\Vancouver,  Canada\\
E-mail: arz@phas.ubc.ca}

\begin{abstract}
We conjecture that  the confinement- deconfinement phase transition in QCD at large number of colors $N$ and $N_f\ll N$ at $T\neq 0$ and $\mu\neq 0$ is triggered by the  drastic change in $\theta$ behavior. 
The conjecture is motivated by   the holographic model of QCD  where 
confinement -deconfinement phase transition 
 indeed happens precisely at   $T=T_c$ where $\theta$ dependence 
experiences a sudden change in behavior. The conjecture is  also supported by 
   quantum field theory arguments  when the instanton   calculations  (which trigger the $\theta$ dependence) are under complete theoretical control for  $T>T_c$, suddenly break down  immediately   below $T<T_c$ with sharp changes in  the $\theta$ dependence.
       Finally, the conjecture is supported by a number of numerical lattice results. We employ this conjecture to study  confinement -deconfinement phase transition of hot and dense QCD   in large $N$ limit  by analyzing the $\theta$ dependence.  We estimate the critical values for $T_c$ and $\mu_c$ 
       where the phase transition happens by approaching the critical values from the hot and/or dense regions where the instanton calculations are under complete theoretical control.
    We also  describe  some defects of various codimensions
within a holographic model of QCD by focusing   on their role
around  the phase transition point.
       
       \end{abstract}


\bodymatter

\section{ Introduction}  
This talk is based on series of recent papers 
   \cite{Parnachev:2008fy,Zhitnitsky:2008ha,Gorsky:2009me}.  
 Understanding the phase diagram at nonzero external parameters $T, \mu$   is one of the most difficult problem in QCD. Obviously, this area  is a prerogative of numerical lattice computations.
 However, some   insights  about  the basic features of the phase diagram may  be inferred    by 
 using some analytical approaches. In particular, some qualitative questions  can be 
formulated and answered by considering a theory with 
  large number of colors $N$. 
We conjecture that  the confinement- deconfinement phase transition in QCD at large number of colors $N$ and $N_f\ll N$ at $T\neq 0$ and $\mu\neq 0$ is triggered by the  drastic change in $\theta$ behavior. 
 This criteria is motivated by   the observation that in holographic model of QCD  the 
confinement -deconfinement phase transition 
 happens precisely at the value of temperature $T=T_c$ where $\theta$ dependence 
experiences a sudden change in behavior\cite{Parnachev:2008fy,Bergman:2006xn}. Secondly,  the proposal  is supported by  the numerical lattice results, see e.g.   review article \cite{Vicari:2008jw},  which unambiguously
suggest that the topological fluctuations are strongly suppressed in deconfined phase, and this suppression becomes more severe with increasing $N$.  Finally, 
 our new criteria is based on a physical picture which  can be shortly summarized as follows.
   
  For sufficiently high temperatures $T>T_c$ the instanton gas is dilute 
with density $\sim e^{-\gamma(T)N}$ which implies  a  strong suppression  of the topological fluctuations
at large $N$ where $\gamma(T)>0$, see below for details on structure of $\gamma(T)-$ function. The calculations in this region are
under complete   theoretical control and the vacuum energy has a nice
analytic behavior $\sim \cos\theta e^{-\gamma(T)N}$ as function of $\theta$. 
At the critical value of temperature, $T=T_c$ where $\gamma(T)$ changes the sign,
 the instanton expansion 
breaks down and 
one should naturally expect  that at $T=T_c$ there should be  a sharp transition in
$\theta$ behavior as  simple formula $\sim \cos\theta$ can only be valid when
the instanton gas is dilute and semiclassical calculations are justified which is obviously not the case
for $T<T_c$. Therefore, it is naturally to 
 associate sharp changes in $\theta$ behavior  with confinement-deconfinement transition,
just as in the holographic model\cite{Parnachev:2008fy}.

 The plan of the paper is as follows.
I start in  Section 2  by considering the phase transition in hot QCD \cite{Parnachev:2008fy}.   
 In section 3   the same technique is applied  to the  dense QCD. I  argue that 
 the confinement- deconfinement phase transition   happens at very large quark chemical potential $\mu_c\sim \sqrt{N}\Lambda_{QCD}$, where $\mu=\mu_B/N$ is already properly
 scaled quark chemical potential\cite{Zhitnitsky:2008ha}.   Finally, in section 4   some defects of various codimensions are described 
within a holographic model of QCD by focusing   on their role
around  the phase transition point\cite{Gorsky:2009me}.

 \section{Deconfinement  Phase Transition in hot QCD at large $N$.   }
 
 We start with analysis of nonzero temperature\cite{Parnachev:2008fy}
   when  a sharp transition in $\theta$ dependence at the phase transition  is indeed observed  in the holographic model of QCD.  From quantum field theory viewpoint such a transition 
 can be understood   as follows. Instanton calculations are under complete theoretical 
 control in the region $T>T_c$ as the instanton density is parametrically suppressed
 at large $N$ in deconfined region\cite{Parnachev:2008fy},
     \beq
  \label{gamma_N}
 V_{\rm inst}(\theta)\sim e^{-\gamma N} \cos\theta,~~~~ \gamma=\Bigl[\frac{11}{3}
 \ln \left(\frac{\pi T}{\Lambda_{QCD}}\right)-1.86\Bigr].
  \eeq
    It is assumed that a higher order corrections may change the numerical 
   coefficients in $\gamma (T)$, but they do not change the structure of eq. (\ref{gamma_N}).
   The critical temperature is determined by condition $\gamma =0$
where  exponentially small expansion parameter  $ e^{-\gamma N}$ suddenly blows up
and becomes exponentially large.
Numerically,
  it happens at 
   \beq
  \label{T_c_N}
  \gamma=\Bigl[\frac{11}{3}  \ln \left(\frac{\pi T_c}{\Lambda_{QCD}}\right)-1.86\Bigr]=0
 ~~~  \Rightarrow ~~~T_c (N=\infty)\simeq 0.53 \Lambda_{QCD},
  \eeq
  where $ \Lambda_{QCD} $ is defined in the Pauli -Villars scheme. 
Our computations are carried out in the regime where the instanton 
density $\sim \exp(-\gamma N) $ is parametrically suppressed  at  any small but finite $\gamma(T)=\epsilon> 0$ when  $N=\infty$.  From eq. (\ref{gamma_N}) one can obtain the following expression for 
instanton density in vicinity of $T>T_c$, 
   \beq
   \label{T}
   V_{\rm inst}(\theta) \sim \cos\theta \cdot e^{-\alpha N\left(\frac{T-T_c}{T_c}\right)}, ~~~~ 1\gg \left(\frac{T-T_c}{T_c}\right)\gg 1/N.
   \eeq
where $\alpha=   \frac{11}{3}$ and $  T_c (N=\infty)\simeq 0.53 \Lambda_{QCD} $ are estimated at one loop level.
 Such a behavior does  imply that the dilute gas approximation is justified even in close vicinity of $T_c$ as long as $\frac{T-T_c}{T_c}\gg \frac{1}{N}$.     
   Therefore, the $\theta$ dependence, which is sensitive to the 
 topological fluctuations is determined by (\ref{T})  all the way down to the temperatures very 
    close to the phase transition point from above, $T=T_c+ O(1/N)$. 
     The topological susceptibility  
 is order of one for $T<T_c$ in confined phase while it vanishes
 $\sim   e^{-\gamma N}\rightarrow 0$ for $T> T_c$ in deconfined phase.
    Non topological quantum fluctuations on the other hand  could be quite large in this region, but they do not effect the structure of eq. (\ref{T}).
     
There are three  basic reasons for a generic structure
(\ref{gamma_N},\ref{T_c_N},\ref{T})  to emerge:\\
{\bf a.} The presence of the exponentially large ``$T-$ independent"  contribution
( e.g. ~$e^{+1.86 N}$ in eq. (\ref{gamma_N})). This term 
   basically describes  the entropy of the configuration. It is due to a number of contributions such as a 
 number of embedding $SU(2)$ into $SU(N)$ etc;\\
 {\bf b.} The  presence of the ``$T-$ dependent" contribution to  $V_{\rm inst}(\theta)  $ which comes  from  $\int n(\rho) d\rho$ integration.  It   is proportional to the standard factor, 
  \beq 
  \label{rho}
     \left(\frac{\Lambda_{QCD}}{\pi T}\right)^{\frac{11}{3}N}=\exp\Bigl[-\frac{11}{3}N
\cdot \ln \left(\frac{\pi T}{\Lambda_{QCD}}\right)\Bigr].
 \eeq
 {\bf c.} The fermion related contributions such as a  chiral condensate, diquark condensate  or non-vanishing mass term    enter the instanton density as follows $ \sim\la\bar{\psi}\psi\ra^{N_f}\sim e^{N\cdot \left(\kappa\ln  |\la\bar{\psi}\psi\ra|\right)} $. For $\kappa\equiv\frac{N_f}{N}\rightarrow 0$ this term  obviously leads to a sub leading effects $1/N$ in comparison
  with two  main terms in the exponent (\ref{gamma_N}) and will be neglected for the rest of this paper.
   
The crucial element in this  analysis   is that both leading contributions (items 1 and 2 above) have exponential $e^N$ dependence,
and therefore at $N\rightarrow \infty$ for $T>T_c$ the instanton gas is dilute with density
$e^{-\gamma N}, ~\gamma>0$ which ensures  a  nice $\cos\theta$ dependence (\ref{T}), while  for $T<T_c$ 
the expansion breaks down, and $\theta$ dependence must sharply change at $T<T_c$.  We have identified such sharp changes with first order phase transition.

One more comment on this proposal. Our  conjecture (that  the confinement- deconfinement phase transition in QCD  is triggered by the  drastic change in $\theta$ at the same point $T=T_c$)   implicitly implies that the configurations  which are responsible for sharp  $\theta$ changes
   must also  drastically change their properties at $T=T_c$.  For $T>T_c$ the objects
   which describe the $\theta$ behavior are the instantons, while at $T<T_c$  we believe, they are the instanton quarks,   the quantum objects with fractional topological charges $\pm 1/N$. Therefore, we  interpret  this transition as dissociation of the instanton into 
  instanton-quarks which become the dominant quasi-particles at $T<T_c$, see more comments on this interpretation in concluding section.

 \section{  Phase Transition  in dense QCD  at  large N.  }
 
 In this section we estimate the value of $\mu_c$ where the instanton expansion breaks
down and therefore, the $\theta$ dependence should experience a sharp change. According to our conjecture we should
identify this position  with the phase transition point.
Similar arguments have been put forward  previously \cite{Toublan:2005tn} for numerical estimation of  $\mu_c$ for small $N, N_f=2, 3$.
Our goal here is quite different: we want to understand an analytical dependence of  $\mu_c (N, N_f)$ 
as a function of $N, N_f$ at very large $N$ and finite $N_f\ll N$ in order to compare 
  with results   of ref.
 \cite{McLerran:2007qj}  where the authors presented  a very strong argument  suggesting a very  large $\mu_c\sim \sqrt{N}$ where the phase transition could happen.
 
  We follow the same logic as before, and study the $\theta$ dependence in
  order to make a prediction about the phase transition point $\mu_c$.
  In the regime $\mu > \mu_c $ the $\theta$ dependence 
 is determined by the dilute instanton gas approximation. We expect that the expansion breaks down only  in close vicinity  of $\mu_c$ at large $N$ as it happens in our previous analysis with phase transition at $T=T_c$. According to the conjecture this point will be identified with 
 confinement- deconfinement  phase transition point $\mu_c$.
 In the present case of  analyzing $\mu_c$ rather than $T_c$ discussed previously   we do not have any support from the  lattice computations, nor from holographic models.
 Still, the basic governing principle remains the same. Therefore we identify    the point where instanton 
 expansion breaks down (and correspondingly a point where  a simple $\cos\theta$ sharply changes to
  something else) with the point $\mu_c$ where the phase transition happens. 
   In our estimates below 
 we assume that the color superconducting phase is realized in deconfined phase
for all $N$, see e.g. recent review \cite{Alford:2007xm}.

As we shall see below, the instanton density    in deconfined phase has the following generic behavior,  $\sim \cos\theta\exp{[-N\gamma (\mu) ]}$, where $\gamma(\mu)\sim const.+0(1/N)$ in large $N$ limit.
 Such a behavior 
implies that for any small (but finite) positive $\gamma>0$ the instanton density 
is exponentially suppressed   
and our calculations are under complete theoretical control.
In contrast: at arbitrary small and negative 
  $\gamma<0$ the instanton expansion obviously breaks down, 
 theoretical control is lost
  as an exponential growth   $\sim \exp{(|\gamma| N)}$ for the instanton density makes no sense.
The  $\theta $ behavior must  drastically change at this point. Therefore, the value of $\mu_c$ is  determined by the following condition, 
  \beq
 \label{mu_c}
 \gamma (\mu=\mu_c) =0 ~~~~~  \Longrightarrow  ~~~~~\mu_c= c\Lambda_{QCD}.
 \eeq
 Our goal is to compute the coefficient $c$ by approaching the critical point $\mu_c$ from deconfined side of phase boundary.
 Therefore,   we will be interested in the instanton density in 
the dilute gas regime at $\mu> \mu_c$  where analytical instanton calculations are under control.
 
    As is well known,   the $\theta$ dependence   goes away in full QCD in both phases: confined as well as deconfined in the presence of the massless chiral fermions.
 However we are interested in the magnitude   of the instanton contribution $\sim V_{\rm inst}(\theta)$     
  in deconfined phase rather than in $\theta$ dependence of full QCD. Precisely this coefficient   triggers  the point where the instanton expansion  suddenly blows up. The sharp changes in $ V_{\rm inst}(\theta)$ we identify with complete reconstruction  of the ground state, drastic   changes of the relevant gluon configurations, and finally, with   confinement- deconfinement phase transition. To avoid identical vanishing of $ V_{\rm inst}(\theta)$ in the presence of massless fermions one can assume 
  a non zero chiral condensate in deconfined phase or  one can   assume a  non-vanishing masses $m_q\neq 0$ for the fermions, or 
non-vanishing diquark condensate $\la{\psi}\psi\ra\neq 0$ to avoid identical vanishing 
of $ V_{\rm inst}(\theta)$ for dense matter at large $\mu$. None of these assumptions
 effects any numerical estimates given below in the limit $N\rightarrow \infty,~ N_f\ll N$,
 as explained in the previous section.
    
   The result of the computations can be represented as follows, 
   \beq
  \label{gamma_mu}
 V_{\rm inst}(\theta)\sim e^{-\gamma N} \cos\theta,~~~~ \gamma=\Bigl[\frac{11}{6}
 \ln \left(\frac{N_f \bar{\mu}^2}{\Lambda_{QCD}^2}\right)-1.1\Bigr], ~~~\mu^2\equiv N\bar{\mu}^2,
  \eeq
  where we introduced reduced chemical potential  $  \bar{\mu}\equiv \mu/\sqrt{N}$ and neglected all powers $N^p$ in front of $ e^{-\gamma N}$.    The crucial  difference in comparison with similar computation
  at nonzero temperature (\ref{gamma_N})  is emerging  of parameter $\bar{\mu}$
  instead of the original quark chemical potential $\mu\equiv \sqrt{N}\bar{\mu}$.
  It implies that the critical chemical potential where $\gamma$ changes the sign
  (and therefore where the phase transition is expected) 
   is parametrically large $\mu_c\sim \sqrt{N}$ because $\bar{\mu}_c\sim 1$, see below
  for numerical estimates. The origin for this phenomenon can be understood from the following observation: the temperature -dependent factor in the instanton density before integrating over $\rho$  is proportional to $\sim N$ while chemical potential enters this expression with factor $\sim N_f\ll N$, see e.g. the review paper \cite{shuryak_rev} . Therefore, a very  large chemical potential $\mu\sim \sqrt{N}\Lambda_{QCD}$ is required in order to achieve the same effect 
  as   temperature $T\sim \Lambda_{QCD}$.     The physics   of this phenomenon can be explained as follows: at $T\sim 1$ a large number of gluons $\sim N^2$ can get excited
  while at $\mu\sim 1$ only a relatively small number of quarks in fundamental representation $\sim N$ 
  can get excited.  Therefore, it requires a very large chemical potential $\mu^2\sim  {N}$ in order  for fundamental quarks play the same role as gluons do at $T\sim 1$.
  As explained above, the critical chemical potential is determined by condition $\gamma =0$
where  exponentially small expansion parameter  $ e^{-\gamma N}$ at $\mu> \mu_c$ suddenly blows up at $\mu<\mu_c$.
Numerically,
  it happens at 
   \beq
  \label{mu_N}
 \gamma=\Bigl[\frac{11}{6}
 \ln \left(\frac{N_f \bar{\mu}^2}{\Lambda_{QCD}^2}\right)-1.1\Bigr]=0
~ \Rightarrow  ~
 \mu_c (N=\infty)\simeq 1.4\cdot \Lambda_{QCD} \sqrt{\frac{N }{N_f}},
  \eeq
  where $ \Lambda_{QCD} $ is defined in the Pauli -Villars scheme. 
 The topological susceptibility  vanishes
 $\sim   e^{-\gamma N}\rightarrow 0$ for $\mu> \mu_c$ while 
 it must be drastically different for $\mu<\mu_c$ as $\theta$ dependence
 must experience some drastic changes in this region as the instanton expansion breaks down, and therefore simple 
 $\cos\theta$ dependence must be replaced by something else.
 It is very likely that the standard Witten's arguments (valid for the confined phase) 
 still hold  in this region $\mu<\mu_c$ in which case the topological susceptibility
  is order of one. 
 
 The $ \Lambda_{QCD} $ in the Pauli -Villars scheme which enters our formula (\ref{mu_N})    is not well-known numerically.
Therefore, for numerical estimates one can trade $ \Lambda_{QCD} $ in favor of $T_c (N=\infty) $ at $\mu=0$ estimated above (\ref{T_c_N}). Therefore, our final numerical estimate for $\mu_c(N=\infty)$ can be presented as
follows, 
  \beq
  \label{mu-final}
 \mu_c (N=\infty)\simeq 2.6\cdot \sqrt{\frac{N }{N_f}}\cdot T_c (N=\infty, \mu=0) , ~~~~ N_f\ll N.
  \eeq
  If one uses the numerical value for $T_c(N=3)\simeq 260$~MeV  
  \cite{Lucini:2003zr,Lucini:2005vg}, one arrives to $\mu_c (N=\infty)\simeq 690  \sqrt{N/N_f}$~MeV  which is our final numerical estimate for the critical chemical potential where deconfined phase transition is predicted for very large $N$. 
Few remarks are in order:\\
{\bf a.}
The most important result of the present studies is the observation that the confinement- deconfinement 
phase transition according to (\ref{mu_N}) happens at very large $\mu_c\sim \sqrt{N}$ if $N_f\ll N$.
This is consistent  with the results of \cite{McLerran:2007qj} where parametrically large scale for $\mu_c\sim \sqrt{N}$ had been predicted.
However, the technique of ref. \cite{McLerran:2007qj} does not  allow to 
answer the question whether the transition would be  the first order   or  it would be  a  crossover.  
Within our framework at $N\gg 1$ and $ N_f\ll N$ the entire   phase transition line  
(which starts at $T=T_c \sim \Lambda_{QCD}$ at $\mu=0$ 
and ends at $\mu=\mu_c\sim \sqrt{N} \Lambda_{QCD}$ at $T=0$)
is predicted to be the first  order phase transition at large $N$ and $N_f\ll N$. This is because the nature  for the phase transition along the entire line  is one and the same: it is drastic changes of $\theta$ dependence when the phase transition line is crossed.\\
{\bf b.} Our computations are carried out in the regime where the instanton 
density $\sim \exp(-\gamma N) $ is parametrically suppressed at $N=\infty$.
From eq. (\ref{gamma_mu}) one can obtain the following expression for 
instanton density in vicinity of $\mu>\mu_c$, 
   \beq
   \label{mu1}
  V_{\rm inst}(\theta) \sim \cos\theta \cdot e^{-\alpha N \left(\frac{\mu-\mu_c}{\mu_c}\right)}, ~~~~ \frac{1}{N}\ll\left(\frac{\mu-\mu_c}{\mu_c}\right)\ll 1,
   \eeq
where $\alpha$ is $11/3$ at one loop level, but the perturbative corrections could be large 
and they may considerably change this numerical coefficient.  
 Such a behavior (\ref{mu1}) does  imply that the dilute gas approximation is justified even in close vicinity of $\mu_c$ as long as $\frac{\mu-\mu_c}{\mu_c}\gg \frac{1}{N}$.    In this case the diluteness parameter  remains small. We can not rule out, of course, the possibility that the  perturbative 
   corrections may change our numerical estimate for $\mu_c$. 
However, we   expect that a qualitative picture of the phase transition advocated in this 
paper remains unaffected  as a result of these   perturbative  corrections   in dilute gas regime.\\
   {\bf c.}  Once $\mu_c$ is fixed   one can compute the entire segment of the  phase transition line $\mu_c(T)$  for    relatively small $T$. 
   Indeed, in the dilute gas  regime at $\mu>\mu_c$  the $T$ dependence of the instanton density is determined
by a simple insertion   $\sim \exp[- 2/3 N \pi^2T^2\rho^2]$ in the expression
for the density, see e.g. the review paper\cite{shuryak_rev} . The result in the leading loop order  $\mu_c(T)$ can be presented  as follows,
\beq
\label{T1}
\mu_c(T)=\mu_c(T=0)\Bigl[1- \frac{N \pi^2T^2}{3N_f \mu_c^2(T=0)} \Bigr], ~~~~~~~~~~~
 \sqrt{N}T\ll \mu_c. 
\eeq
\exclude{One should remark that a variation of the critical chemical potential $\Delta \mu_c(T)$ is very large  $\sim \sqrt{N}$ when the temperature variation   $\Delta T\sim 1$ is  order of one in units of $\Lambda_{QCD}$. This is in huge contrast with a similar expression for hot QCD which shows very little change  $\sim 1/N$ of the critical temperature $\Delta T_c(\mu)\sim 1/N$  with   variation of chemical potential of order one, $\Delta \mu\sim 1 $. 
 The nature    of this difference between $\mu_c$ and $T_c$  was already mentioned before and can be explained by  the fact  that at  $T\sim 1$ a large number of gluons $\sim N^2$ can get excited
  while at $\mu\sim 1$ only a relatively small number of quarks in fundamental representation $\sim N$ 
  can get excited.  Therefore, it requires a very large chemical potential $\mu^2\sim  {N}$ when quarks can   play the same role  as gluons do at  $T\sim 1$ as long as $N_f \ll N$.}

 \section{QCD defects via holography}
 Our discussions in the previous sections were based on the instanton calculus in vicinity of the phase transition. The instanton in the holographic description can be  identified with the D0 brane extended along $x_4$,  see  refs \cite{Parnachev:2008fy, Gorsky:2009me,Bergman:2006xn} for notations and introduction into the subject. The holographic picture gives a very natural way to represent the deconfinement phase transition as the Hawking-Page phase transition   in which case   the two metrics with the same asymptotics get interchanged. As our main goal in this talk is   to understand the phase transition, it would be interesting to study other gauge configurations from 
 the holographic viewpoint.  In what follows we limit ourselves only
 with two examples. For a    more complete list of different defects which can be studied by this technique, see the original paper\cite{Gorsky:2009me}.
 
{\it \underline{D4 particle}}\\
There are several possible embeddings of D4 branes.  One possibility
is to wrap   D4 brane  around $S^4$ and extended along the Euclidean time $\tau$, see 
\cite{Gorsky:2009me} for the details. 
The key point here is that  due to this wrapping around $S^4$
the ``electric charge" $N$ is induced
on the  D4 brane.  The resulting 3 dimensional system 
  is  a static and topologically stable  configuration. 
If we had quarks we could make a baryon by attaching quarks to open strings.
In the pure YM case there are no flavor branes describing quarks  to make a gauge invariant object.
 The most simple way to achieve this  goal is to add
$N$ D0 branes yielding the D0-D4 open strings. Hence we get the D4 particle
which is not sensitive to the  $\theta$-term. It is important that this configuration   is well-defined below the critical temperature $T<T_c$ in contrast with instantons discussed in 
\cite{Parnachev:2008fy,Bergman:2006xn} when D0 branes were sensitive to the $\theta$ term and were well defined above the critical temperature $T>T_c$. 
\begin{figure}
\epsfxsize=10cm
\centerline{\epsfbox{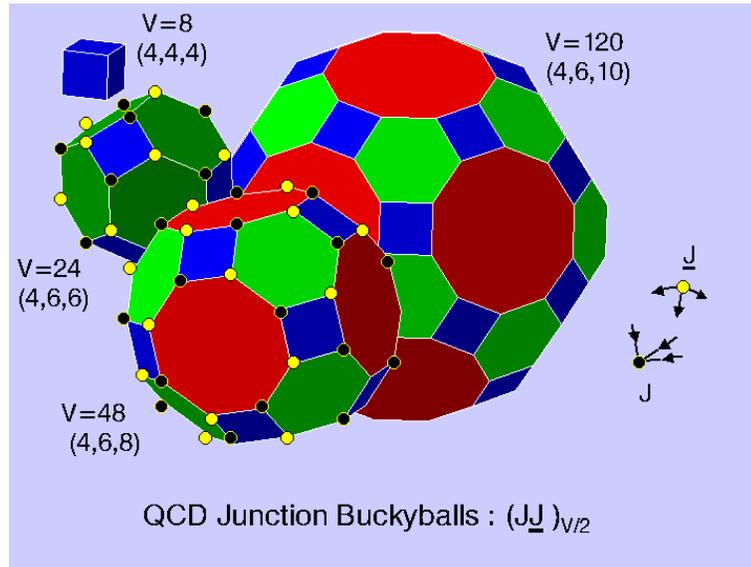}}
\caption{\label{Fig1}
 Buckyballs with $N=3$ and $V=8, 24, 48, 120$ from \cite{Csorgo:2001sq}.
The construction combines  `` D4 particle"
with  ``D4 anti-particle" to form a gauge invariant object with zero baryon charge. The mass of this object scales as $\sim N$. }
\end{figure}
We can  combine the `` D4 particle"
with  ``D4 anti-particle" to form a gauge invariant object, see Fig 1. 
The mass of this object scales as $\sim N$ and is much heavier  than the usual glueballs. In fact, one can construct
gauge invariant objects with any even number of vertexes such that the total charge vanishes.
It is a new family of glueballs with mass $\sim N V$, where $V$ is the number of D4 and anti-D4 particles which form a desired configuration.  
It is amusing that such kind of structure in QCD had been previously discussed \cite{Csorgo:2001sq}
motivated by the discovery of the carbonic Fullerenes $C_{60}$ and $C_{70}$ in 1985, 
which are nano-scale objects \cite{fullerenes}.  
The QCD objects, similar to the carbonic Fullerenes with femto-meter scale were named Buckyballs. 
It has been also demonstrated that the ``magic" numbers for  Buckyballs are $V=8, 24, 48, 120$ 
which correspond to the most symmetric, and likely, most stable configurations \cite{Csorgo:2001sq}. 
The properties of this configuration are not sensitive to $\theta$. 
The possibility to discover such kind of configurations at RHIC was discussed in  
\cite{Csorgo:2001sq}.
\\

 {\it\underline{D0-D2}.}\\
In this subsection we want to address the following question: what happens to  instantons (represented by D0 branes in holographic description) 
 in the confined phase?
  Naively, one could think that   the system becomes unstable at $T<T_c$, and therefore the instantons simply disappear from the system. However,   one should speak about effectively zero action for formation of such kind of objects rather than about their instability. 
 Therefore, numerous number of these objects can emerge  in the system without any suppression.  
In fact,  it has been argued in \cite{Parnachev:2008fy} that this is precisely what is happening when
 one crosses the phase transition line from above.

It turns out that there are configurations in holographic picture which exactly describe dissociation 
of the instanton into N different objects with fractional topological charge $1/N$. The construction is described in \cite{Gorsky:2009me} and requires 
N  additional   D2 domain walls   localized at some points along
$x_4$ coordinate. In this construction  an object with a fractional topological charge $1/N$ may emerge. 
Indeed, one can follow the construction of ref. \cite{Davies:1999uw} for SUSY case
when  $N$ D2 branes  located symmetrically split the instanton into $N$
constituents stretched between pairs of
domain walls. Each constituent has fractional instanton number $1/N$  as well as the
fractional monopole number and has no reason to condense.
In our system we have precisely appropriate D2 branes which are needed for this construction.
These monopoles are instantons in the 3d gauge theory on the
D2 worldvolume theory which involves the  scalar corresponding to the position
of D2 branes. As we speculate   below  these magnetic monopoles
may play an important role in the region close to the phase transition $0 < |T-T_c|/T_c\leq 1/N$.

\section{Conclusion. Speculations.}

A  general  comment on this proposal can be formulated as follows.
 Our conjecture which relates two apparently unrelated phenomena (phase transition vs sharp changes in $\theta$ behavior)   implicitly implies that topological configurations which are linked to
 $\theta$ must play a crucial role in the dynamics of the phase transition. For $T>T_c$ such configurations are well-known: they are dilute instantons with density $\sim e^{-\gamma N}\cos\theta$.
 We presented arguments in\cite{Parnachev:2008fy} 
(see also earlier references therein) suggesting that at $T<T_c$ the instantons do not disappear from the system, but rather dissociate into  fractionally charged   constituents, the so-called instanton quarks.   In this sense the phase transition can be understood as a phase transition between molecular phase (deconfined)
  and plasma phase (confined) of these fractionally charged   constituents.  
   A similar  conclusion on sharp changes in $\theta$ behavior at $T=T_c$   was also observed  in ref.\cite{Gorsky:2007bi} where
   the authors studied the D2 branes in confined and deconfined phases at $T\neq 0$.
 The topological objects (sensitive to $\theta$)
were identified  in ref.\cite{Gorsky:2007bi} as magnetic strings. 
  
 If the picture advocated in the present work about the nature of the transition turns out to be correct, it would strongly suggest  that fractionally charged   constituents (which carry the magnetic charges as discussed in\cite{Parnachev:2008fy}) may play a very important
  role  in dynamics in deconfined phase in close vicinity of the  transition $0< (T-T_c)\leq 1/N$. 
  For large $N$ 
  this region shrinks to a point, however for finite $N$ it could be an extended region in temperatures.
  In this region the instantons are not formed yet, and our semiclassical analysis is not justified yet
  as eq. (\ref{T}) suggests. 
  However, the constituents in this region are already  not in condensed form. Therefore they may   become an important  magnetic degrees of freedom which may contribute
  to the equation of state
  similar to analysis on
  wrapped monopoles in ref.\cite{Chernodub:2006gu}.
  \exclude{
  The role of these fractional magnetic constituents could be even more profound if  $N_f\sim N$ 
 where it is known that smooth crossover rather than phase transition likely to take place.  In this    case the region of interests is    
 order of one  $(T-T_c)\sim 1$ in $\Lambda_{QCD} $ units in contrast with  a narrow region $0< (T-T_c)\leq 1/N$ if the first order phase transition takes place.    The region above $T_c$ is also very interesting from phenomenological viewpoint  as reviewed in \cite{Shuryak:2008eq}.
}

  To conclude  this talk I  want to make a few comments on two recent papers  \cite{Unsal:2008ch,Diakonov:2007nv} where the authors advocate the picture similar to the one presented here.

I start with \cite{Unsal:2008ch}.  In that  work the authors consider a 
specifically deformed $SU(N)$ gluodynamics at $T\neq 0$.
It has been shown that such a deformation  
supports a  reliable analysis  in the weak coupling regime in the confining phase. The results of the corresponding calculations imply
that the relevant degrees of freedom in the confined phase are the 
self dual magnetic monopoles with action $\frac{8\pi^2}{g^2N}$
and  topological charges $Q=\pm 1/N$ which are precisely the features of the instanton quarks 
discussed above. 

In contrast, the starting point of ref.\cite{Diakonov:2007nv} is  semiclassical calculations in the background  of calorons  where the weak coupling regime can not be guaranteed.  While the calculations are semiclassical in nature, and therefore, can not be trusted in the strong coupling regime, still, the corresponding analysis   
shows how well localized instantons with integer topological charges at $T>T_c$  
may dissociate into the fractional constituents at $T<T_c$, and become the key players
in the confining phase. This is precisely the picture we are advocating in the present work
based on analysis of sharp $\theta$ changes at $N=\infty$.  It is impressive how
complicated  semiclassical calculations carried out in \cite{Diakonov:2007nv} 
lead to the expression for the vacuum energy $E_{vac}\cos(\frac{\theta}{N})$ advocated in \cite{HZ}
using completely different technique. 
 
  Our technique does not allow us to make any dynamical calculations in this phase 
  as all color degrees of freedom  have been integrated out in obtaining the low energy effective lagrangian. In other 
words, we can not study the dynamics   
 of fractionally charged constituents in contrast with  papers \cite{Unsal:2008ch, Diakonov:2007nv}
 \footnote{In particular, we do not see a beautiful picture of a multi-component   color  Coulomb 
plasma with nearest-neighbor   interactions in the Dynkin space advocated in \cite{Unsal:2008ch, Diakonov:2007nv}. 
 Still, we do see the color- singlet Coulomb interaction of the fractionally   charged   $\pm 1/N$ constituents due to 
$\eta'$ at very large distances where color already confined. }. 
 However, the fact that the constituents carry fractional topological charge $1/N$
  can be recovered  in our approach because the color- singlet 
 $\eta'$ field enters the effective lagrangian as    $ \cos(\frac{\theta-\varphi}{N})$ and serves as a perfect probe   of the  topological charges of the constituents. 
One should also emphasize that the procedure of the recovering of the fractional  topological charge $1/N$ (which has been used here) is not based on the weak coupling expansion. 

It was great  pleasure to participate in the Workshop honoring 60th anniversary of Misha Shifman. Happy birthday, Misha!

 This work 
was supported, in part, by the Natural Sciences and Engineering
Research Council of Canada.

\end{document}